\documentclass{jfm}

\usepackage{graphicx}
\usepackage{color}
\usepackage{float}
\usepackage[normalem]{ulem}

\begin{document}

\newtheorem{lemma}{Lemma}
\newtheorem{corollary}{Corollary}

\shorttitle{Water entry of deformable spheres} 
\shortauthor{Hurd et al.} 

\title{Water entry of deformable spheres}

\author
 {
Randy C. Hurd\aff{1}
  \corresp{\email{randyhurd@gmail.com and taddtruscott@gmail.com}, splashlab.org},
  Jesse Belden \aff{2}
  Michael A. Jandron \aff{2}
  D. Tate Fanning \aff{3} 
  Allan F. Bower \aff{4}
  \&
  Tadd T. Truscott \aff{1}
  }

\affiliation
{
\aff{1}
Dept. of Mechanical \& Aerospace Engineering, Utah State Univ., Logan, UT 84321, USA
\aff{2}
Naval Undersea Warfare Center Division Newport, Newport, RI 02841, USA
\aff{3}
Dept. of Mechanical Engineering, Brigham Young Univ., Provo, UT 84602, USA
\aff{4}
School of Engineering, Brown Univ., Providence, RI 02912, USA
}

\maketitle

\begin{abstract}
When a rigid body collides with a liquid surface with sufficient velocity, it creates a splash curtain above the surface and entrains air behind the sphere, creating a cavity below the surface. While cavity dynamics have been studied for over a century, this work focuses on the water entry characteristics of deformable elastomeric spheres, which has not been studied. Upon free surface impact, elastomeric sphere deform significantly, resulting in large-scale material oscillations within the sphere, resulting in unique nested cavities. We study these phenomena experimentally with high speed imaging and image processing techniques. The water entry behavior of deformable spheres differs from rigid spheres because of the pronounced deformation caused at impact as well as the subsequent material vibration. Our results show that this deformation and vibration can be predicted from material properties and impact conditions. Additionally, by accounting for the sphere deformation in an effective diameter term, we recover previously reported characteristics for time to cavity pinch-off and hydrodynamic force coefficients for rigid spheres. Our results also show that velocity change over the first oscillation period scales with a dimensionless ratio of material shear modulus to impact hydrodynamic pressure. Therefore we are able to describe the water entry characteristics of deformable spheres in terms of material properties and impact conditions.
\end{abstract}


\section{Introduction}
Water entry has been studied for over 100 years, with the earliest images taken by Worthington at the turn of the century \citep{Worthington1908}, and much of the foundational work performed in the 1950s and 60s with military application in mind \citep{Richardson1948, May1948, May1952}. The topic of water entry is still of interest today with several significant research papers published in the last 20 years, investigating topics such as cavity physics, projectile dynamics and even ricochet off the water surface (\cite{Truscott2014,Aristoff2009,Duez2007,Duclaux2007,Seddon2006,Belden2016}, respectively). 

\begin{figure}
  \centering
  \includegraphics[width=4in]{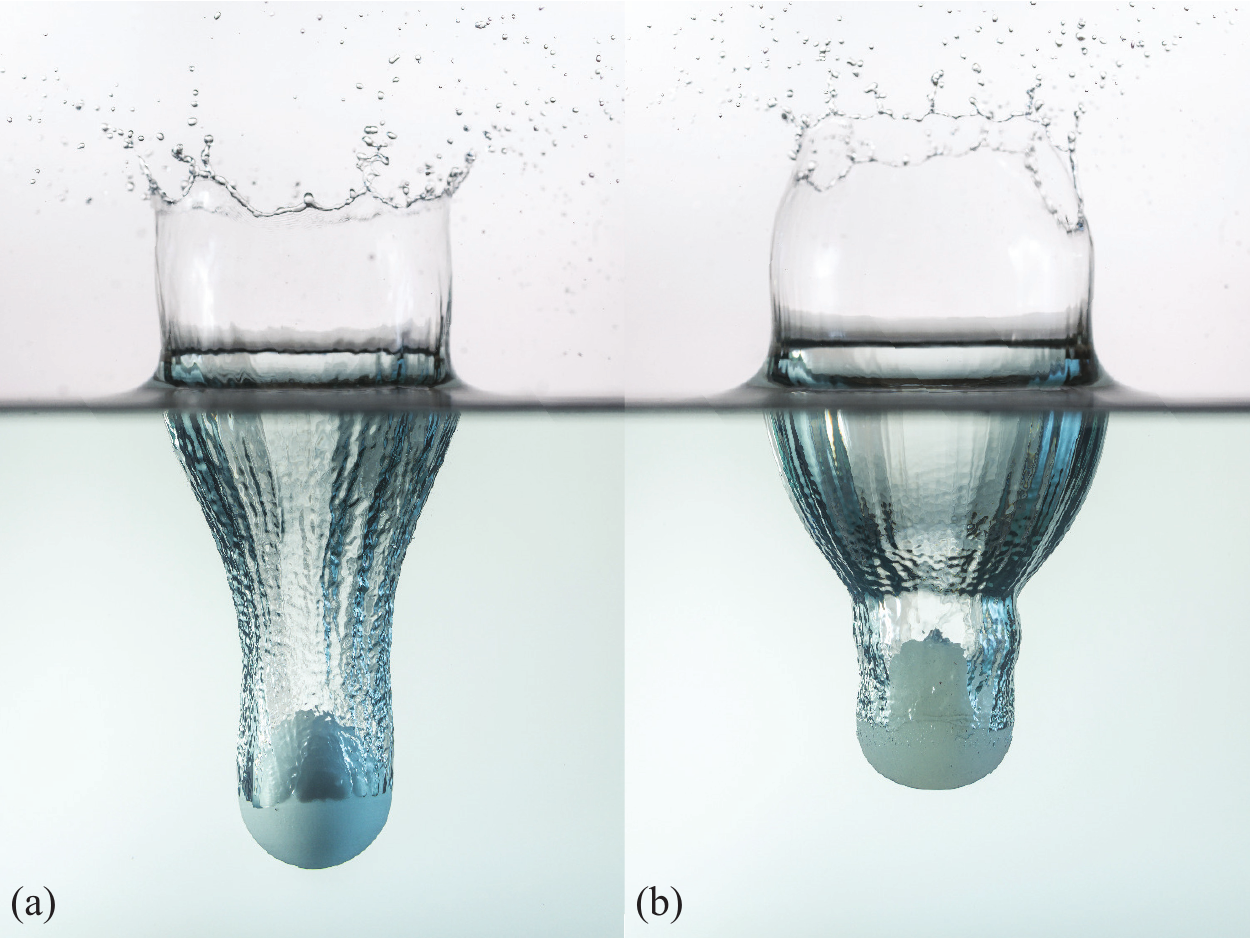}
  \caption[parameters]{ 
Two spheres with shear moduli $G_{\infty}$ differing by four orders of magnitude, experience very different water entry dynamics. (a) A rigid sphere ($G_{\infty}$ = 5.66 $\times 10^5$  kPa), with a solid-liquid density ratio near unity, impacts the free surface forming a canonical subsurface cavity. (b) A deformable sphere ($G_{\infty} = 12.69$ kPa), with otherwise nearly identical properties and impact conditions as (a), forms an altered subsurface cavity due to relatively large deformations and material oscillation. Images were taken at the same time after impact. (Photo credit C. Mabey.)}
\vspace{-0.3cm}
  \label{fig:high_res}
\end{figure}

\vspace{-0.1cm}
Cavity characteristics vary with Froude Number, Bond Number, Capillary Number and by varying object geometry, rotation and wetting angle \citep{Truscott2014}. For example, low speed impact events with sufficiently small capillary numbers will not form subsurface cavities \citep{Duez2007}. When a cavity forms it is often described by the manner in which the cavity collapses (or pinches-off). Cavities are categorized according to the depth at which pinch-off occurs, and these categories include: surface seal, deep seal, shallow seal and quasi-static seal \citep{Aristoff2008, Aristoff2009}. The results herein occur within the high Bond number parameter space ($Bo$ $>$ 300), where surface tension is negligible and only deep seal type pinch-off events have been observed. Previous studies provide theoretical predictions for pinch-off time and depth which produce good agreement with experiments employing steel spheres (high solid-liquid density ratio) where deceleration can be neglected. \citet{Aristoff2010} revealed that a small mass ratio associated with a decelerating sphere can reduce the depth of pinch-off, but does not alter pinch-off time. 

Beyond revealing scaling for pinch-off depth and time, several studies have explored the effect of unique impact conditions on cavity physics and body dynamics. Simply changing the geometry of the projectile generates a cavity with a cross-section resembling the outer profile of the impacting body \citep{Enriquez2012}. For slender-bodies it has been shown that even nose shape and entry angle can greatly alter cavity form and dynamics \citep{May1952, Bodily2014}. Spinning the projectile perpendicular to the free surface prior to impact creates asymmetrical cavities and generates unbalanced forces \citep{Truscott2009b}. Similar findings have resulted from covering half of a hydrophilic sphere with a hydrophobic coating \citep{Truscott2009a}. Both of these methods generate asymmetrical cavities and cause the impacting body to veer from the primary axis of travel. {Some groups have extended the work to biological organisms, for instance, \citet{Chang2016} experimentally investigated plunge-diving birds using a simplified model. Their experiment involved an elastic beam attached to a rigid cone (representing the bird neck and head, respectively), and focused specifically on when buckling occurs as it relates to possible physical damage, as opposed to how significant deformations affect cavity shape and entry dynamics as discussed herein.

Recently, the authors investigated deformable spheres impacting a water surface at an oblique angle, primarily concerned with the effect of deformability on ricochet \citep{Belden2016}. It was shown that induced vibrations interact with the cavity in unique ways resulting in nested cavities, but also inefficient skipping. However, we are not aware of any research addressing the normal entry behavior of deformable elastomeric spheres. Fig.~\ref{fig:high_res} presents two high resolution photographs which qualitatively display some of the differences between the water entry of rigid and highly deformable spheres, including differences in cavity shape and sphere deceleration. In this paper, we use an experimental approach to investigate the unique phenomena associated with water entry of highly deformable spheres.

\vspace{-0.1cm}
\section{Methods} 

We investigated the water entry characteristics of elastomeric spheres experimentally by varying sphere impact velocity $U_0$, diameter $D$ and material stiffness, as characterized by the neo-Hookean shear modulus $G_{\infty}$. Spheres were made from an incompressible platinum-cure silicone rubber called Dragon Skin$^{\tiny{\circledR}}$, which is produced by Smooth-On, Inc. Shear modulus was varied by adding a silicone thinner to the mixture to produce three discrete values  ($G_{\infty}$ = 1.12, 6.70 \& 70.2 kPa), which were determined by sphere compression tests (see Appendix~\ref{appA}). The constituents of the silicone rubber were measured by mass ratio, mixed, then placed in a vacuum chamber to remove entrained air. Mixtures were poured into aluminum molds to form spheres with two different diameters ($D$ = 51 \& 100 mm). Spheres had a density of $\rho_s = 1070$  $\mathrm{kg/m^3}$, and the density of water is represented by $\rho_w$. The water entry of rigid spheres with identical $\rho_s = 1070$ $\mathrm{kg/m^3}$ were also investigated for comparison.

The experimental setup is summarized in Fig.~\ref{fig:setup}a. Spheres were dropped from three discrete heights (0.53, 1.53 $\&$ 2.27 m) into a $0.81 \times 0.81$ m$^2$ glass tank  filled to $\sim$ 1 m with water. The entry event was filmed using two Photron SA3 high speed cameras at 2000 frames per second with diffuse back lighting. The scalar $\lambda$ represents the deviation of the deformed sphere from the initial diameter. Before splash curtain dome over, the changing diameter $\lambda D$ of the sphere was measured by fitting a circle (cyan) to the top view of the sphere as shown in Fig.~\ref{fig:setup}b. After dome over, $\lambda D$ was measured below the free surface (side view Fig.~\ref{fig:setup}c). The lowest point of the sphere $y_b$ (red cross Fig.~\ref{fig:setup}c) was also measured directly from the images. The separation line at the air-water-sphere interface is marked by a green horizontal line. An ellipse was then fitted to the edge of the sphere below the separation line (yellow outline).  Because the sphere deformation is assumed to be symmetric about the $y$-axis, measurements of $\lambda$ from the side and top camera views are assumed to represent the same quantity. 

\begin{figure}
  \centering
  \includegraphics[width=\textwidth]{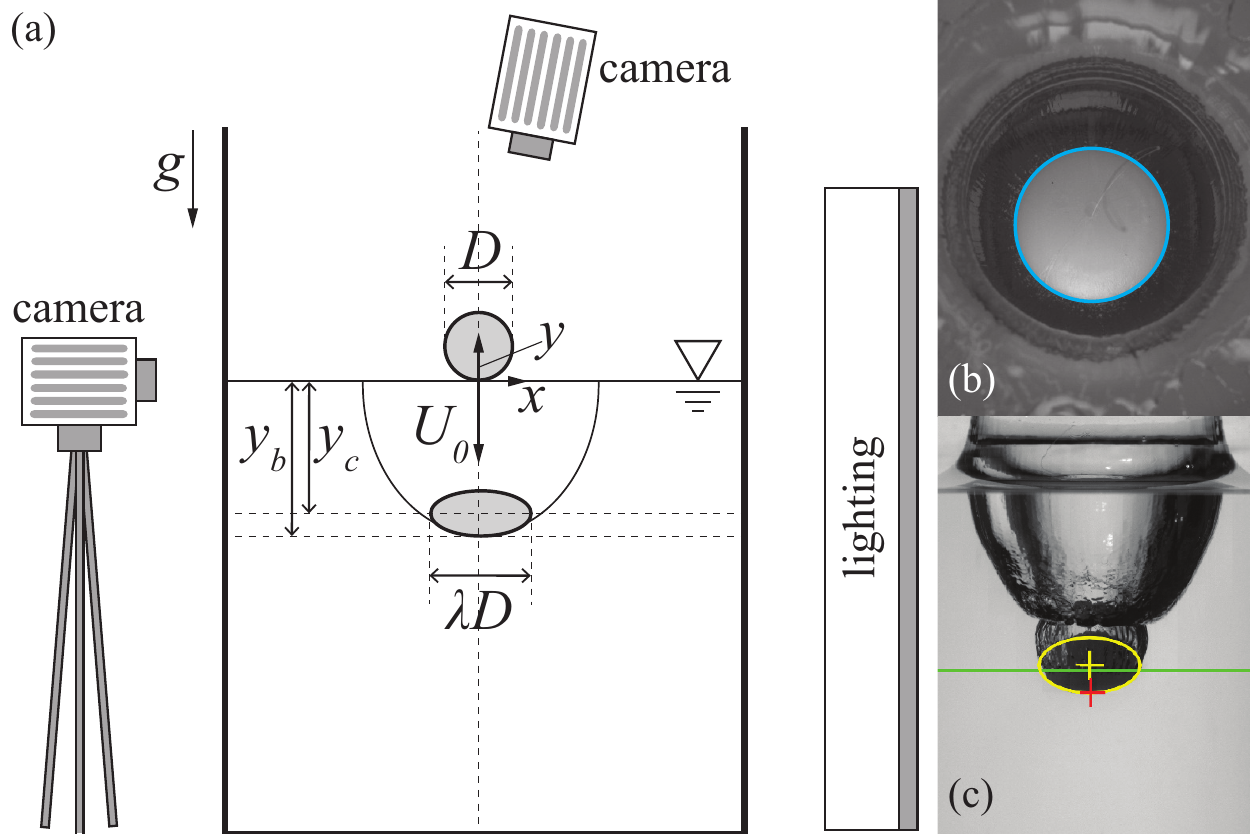}
  \caption[Visualization of image processing]{ 
(a) Spheres falling into a water-filled tank were filmed using high speed cameras and diffuse back lighting. The event is described by the parameters: sphere diameter $D$, impact velocity $U_0$,  a scalar defining sphere deformation $\lambda$, and the depth below the free surface of the sphere's lowest point $y_b$ and the sphere center $y_c$. (b) Image processing was used to measure $\lambda D$ prior to dome over. (c) Image processing was used to locate $y_b$ throughout the entry event (red cross). A curve was fitted to the edge of the sphere below the air-water-sphere interface (green line). A corresponding ellipsoid, with an assumed symmetry about the $y$-axis and a volume constrained by the undeformed sphere, was applied (yellow outline). 
}
\vspace{-0.3cm}
\label{fig:setup}
\end{figure}

\section{Results} 
Fig.~\ref{fig:high_res} displays high resolution images of two spheres with nearly identical impact conditions ($U_0$ = 2.4  m/s, $D$ = 51  mm), except that the sphere in (a) has shear modulus $G_{\infty}$ = $5.66 \times 10^5 $ kPa (rigid) while the sphere in (b) has shear modulus $G_{\infty}$ = 12.7 kPa. The cavity formed by the deformable projectile differs in the oscillatory profile of the cavity walls, in addition to being shallower and wider. Because the cavity physics and projectile dynamics are evidently different for a deformable sphere, characterizing the initial deformation and resulting material oscillation in the sphere is critical to understanding water entry physics for deformable objects. 

Fig.~\ref{fig:exp}a\&b lends additional insight into why a cavity produced by a deformable sphere deviates from that formed by a rigid sphere. At 12 ms after impact, the sphere has deformed significantly into an oblate spheroid, creating a wider cavity than a rigid sphere. Elastic forces cause the sphere to rebound from this initial deformation into a prolate spheroid with its major axis aligned with the vertical ($t$ = 29 ms). The continually oscillating sphere now proceeds to a second radial expansion that penetrates through the cavity wall, forming a smaller cavity within the first ($t$ = 43 ms), resulting in a so-called matryoshka cavity \citep{Belden2016}, \citep{Hurd2015}. For this case pinch-off occurs within the second cavity ($t$ = 96 ms). 

For each experimental test, the position of the bottom of the sphere $y_b$ was tracked through a series of high speed images. In Fig.~\ref{fig:exp}c, $y_b/D$ is plotted as a function of dimensionless time $t/t_p$ ($t_p:$ time to pinch-off), in which the sphere oscillation is evident. Fig.~\ref{fig:exp}d shows the measured value of $\lambda$, which reaches an initial large peak due to the impact event ($t = 12$ ms), and then decays throughout the water entry.  This decay in $\lambda$ is typical for all deformable sphere water entry events studied.

\begin{figure}
  \centering
  \includegraphics[width=\textwidth]{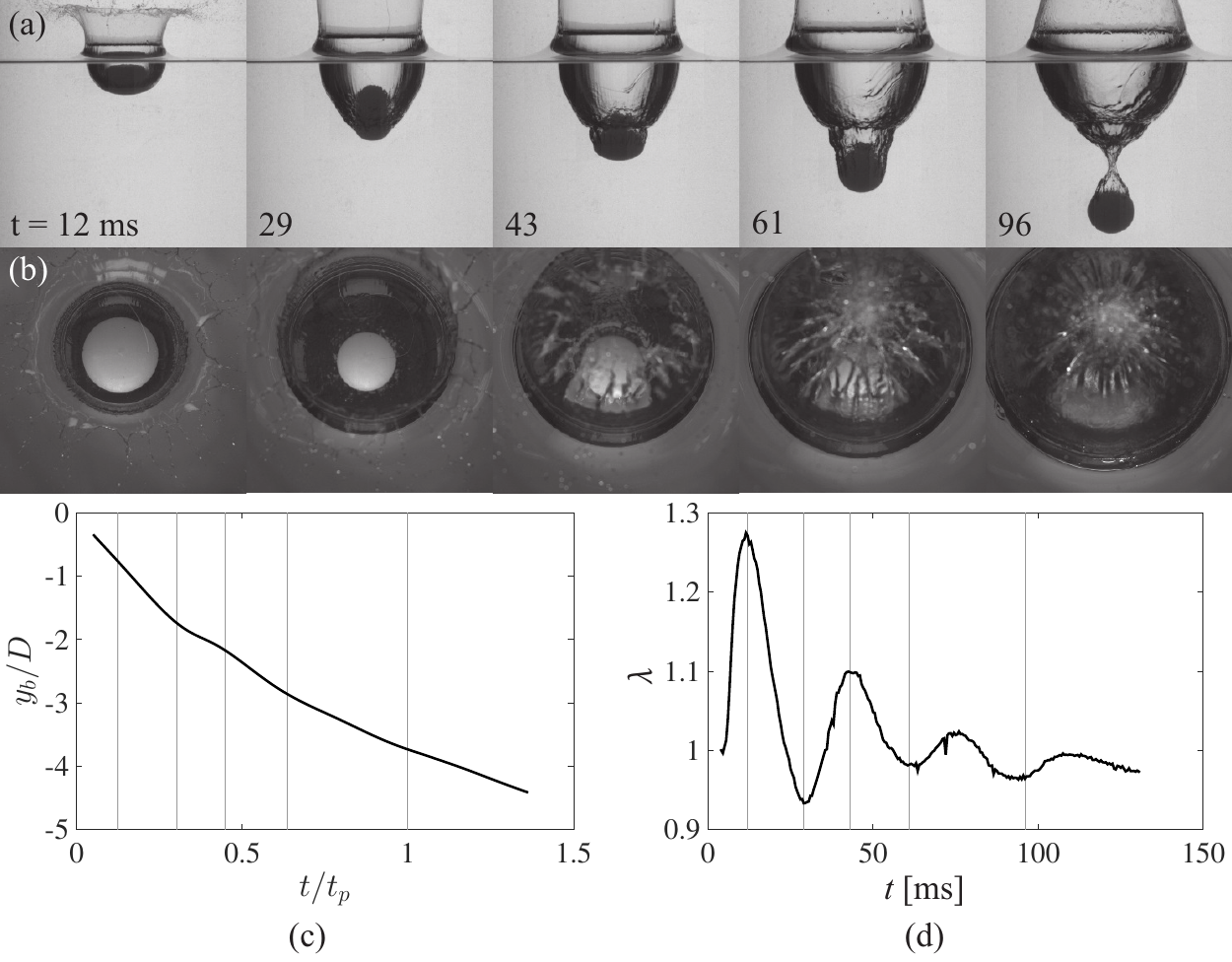}
  \caption[experimental results]{ 
(a) A sphere deforms significantly as it impacts and enters the water ($G_{\infty}$ = 6.70 kPa, $D$ = 51 mm and $U_0$ = 5.3 m/s); after the initial deformation the sphere oscillates between oblate and prolate shapes, creating a second cavity within the first. Pinch-off occurs within this second, smaller cavity ($t=$ 96 ms, supplemental movie 1). (b) The water entry event captured from a  top view highlights the changing diameter and splash curtain dome over event. (c) The measured position of $y_b$ is plotted against dimensionless time where vertical lines correspond to the images above. (d) Plotting the parameter $\lambda$ as a function of time portrays a decaying sinusoid. Image sequences (a) $\&$ (b) are both shown in supplemental movie 1.
}
\vspace{-0.3cm}
  \label{fig:exp}
\end{figure}

Fig.~\ref{fig:lambda_scaling}a presents a simplified description of the sphere oscillation in which the sphere deforms into an oblate spheroid with symmetry about the $y$-axis. Here, $\lambda$ represents the principal stretch in the $x$ and $z$ directions, and by conservation of volume the principal stretch in the $y$ direction is $\lambda_y = 1 / \lambda^2 $. Defining $\lambda$ in this way is based on the observation that the primary mode of deformation in the sphere during water entry is equi-biaxial tension, and $\lambda$ is a measure of the principal stretch in the sphere. The parameter $\lambda_{pN}$ represents the maximum stretch of the sphere in the $x$-$z$ plane for the $N^{th}$ deformation period.

Based on the decaying behaviour of $\lambda$ during water entry, we aim to address if the source of damping is in the sphere material, water or both.  First, we isolate the response of the sphere by performing a series of tests in which the spheres are dropped onto horizontal rigid surfaces (Appendix~\ref{appA}, Fig.~\ref{fig:table_impact}, supplemental movie 2).  Impact with the rigid surface results in an initially large sphere deformation that decays in time. Based on these observations, we apply a viscoelastic model to the sphere material~\citep{Bergstrom1998} as summarized in Appendix~\ref{appA}.  The model includes parameters to account for viscous damping, but the equilibrium stress is still governed by the hyperelastic neo-Hookean model (parameterized by shear modulus $G_{\infty}$). The rigid surface impact data is used to calibrate the dynamic parameters of the viscoelastic model. Second, to see if the damping in the material model can explain the decay observed in the water entry events, we construct a simplified model of the sphere oscillation (derived in Appendix~\ref{appA}).  The sphere is prescribed an initial stretch $\lambda_0 = \lambda_{p1}$ at $t = t_0$ and allowed to oscillate freely for $t>t_0$. 
The analysis is performed for all experimental cases and the results are summarized in Fig.~\ref{fig:lambda_scaling}.

Figure~\ref{fig:lambda_scaling}b shows that the oscillation period predicted by the model is slightly less than that observed in the experiments.  Because the viscoelastic model parameters were calibrated to an experimental test isolated from the water, we suggest the observed lower frequency (longer period) of the sphere in water is attributable to the added mass experienced by the sphere. Some mass of water has to be accelerated during portions of the sphere oscillation period (e.g., between $t=3T/4$ and $t=5T/4$ in Fig.~\ref{fig:lambda_scaling}a).  However, if we scale the response in time by a ratio of experimental to modeled periods, $T_{\mathrm{exp}}/T_{\mathrm{model}}$, then the predicted oscillations show good agreement with the experiments. Despite the difference in period, the magnitude of the predicted peaks in $\lambda$ are consistent with experiments (Fig.~\ref{fig:lambda_scaling}c\&d), suggesting that the dominant source of damping is in fact in the sphere material. We note that the model agrees more accurately at the peaks than the valleys because an oscillating sphere complies with the idealization of the model (ellipsoidal assumption) more closely in its oblate shape than its prolate shape as is observed in Fig.~\ref{fig:allsoft_vs_rigid}d.


\begin{figure}
  \centering
  \includegraphics[width=\textwidth]{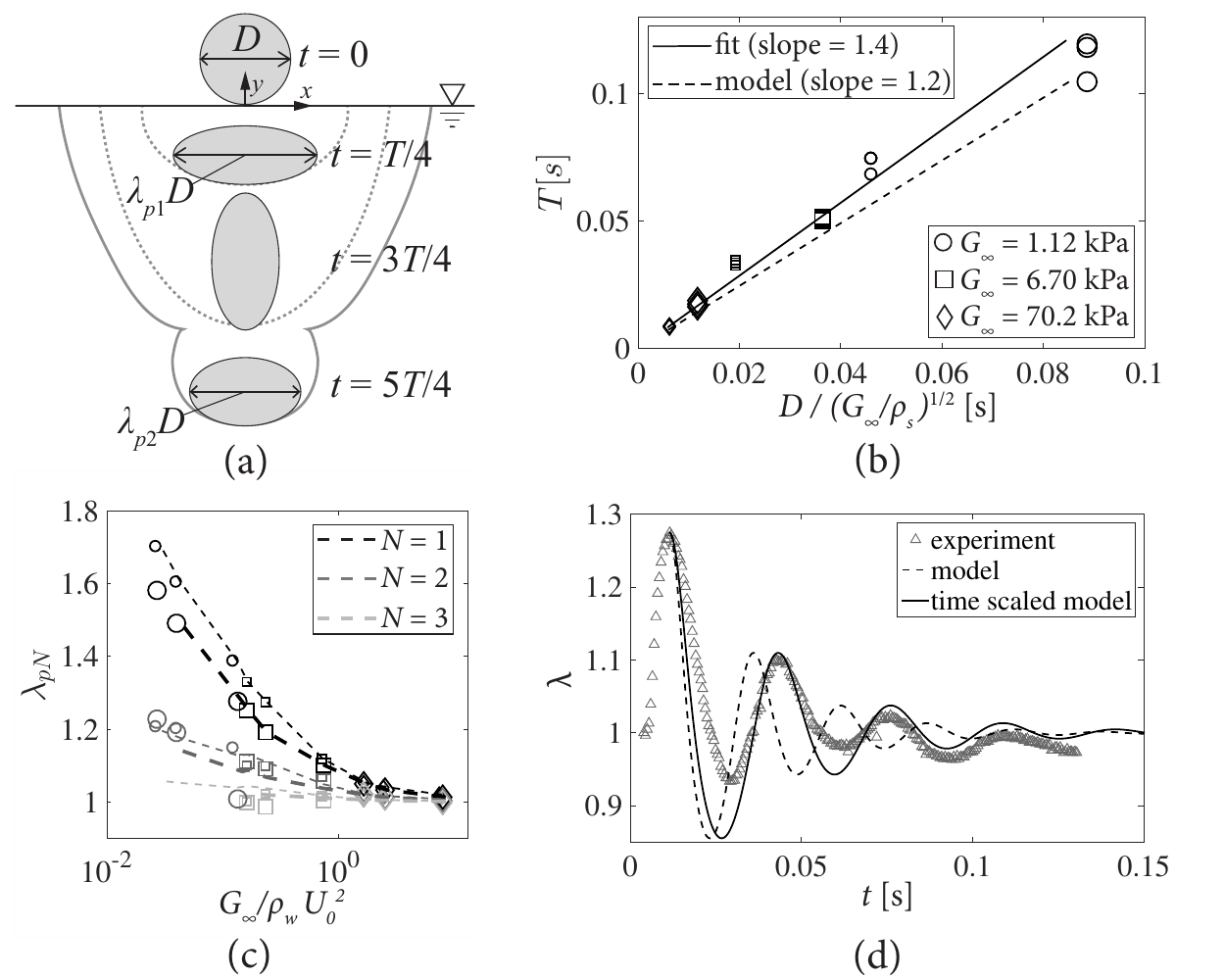}
  \caption[scaling for sphere deformation and vibration]{ 
(a) When a deformable sphere impacts the water surface it flattens into an oblate spheroid with an increased cross-sectional diameter at $t = T/4$ ($\lambda_{p1} D$) before rebounding back into a sphere  at $T/2$. The sphere then forms a prolate spheroid at $3T/4$, then returns to a spherical shape in a single period $T$ (not shown). The principal stretch $\lambda$ defines the deviation of the sphere from its un-stretched diameter $D$. The subscript $pN$ refers to the maximum value of $\lambda$ within the $N^{th}$ period. (b) The oscillation period of $\lambda$ scales with $D / \sqrt{G_{\infty}/\rho_s}$ (the slope of the linear fit is 1.4). The analytical model predicts a slightly smaller slope (1.2). Large data points represent large spheres ($D$ = 100 mm) and small data points represent small spheres ($D$ = 51 mm). (c) The peak value of $\lambda$ for a given period and given sphere radius appears to depend only on $G_{\infty}/\rho_w U_0^2$. Shading denotes period number ($N$) as indicated in the legend. Symbols represent experimental data and lines represent model prediction. Thin lines represent small spheres ($D$ = 51 mm) and thick lines represent large spheres ($D$ = 100 mm). (d) The measured values of $\lambda$ in time are represented by grey triangles. The behavior predicated by the analytical model is represented by a dashed line. The solid line portrays a time-adjusted model with frequency shifted to correspond with the scaling in (b).
}
\vspace{-0.3cm}
  \label{fig:lambda_scaling}
\end{figure}

Fig.~\ref{fig:allsoft_vs_rigid}a-d show the cavity growth and pinch-off resulting from the impact of four spheres ($D$ = 51~mm, $U_0$ = 6.5 m/s) with $G_{\infty}$ decreasing from $5.66 \times 10^5$ kPa (a) to 1.12 kPa (d). The cavity in image sequence (b) is created by a sphere with a shear modulus of $G_{\infty}$~=~70.2~kPa. The resulting cavity and pinch-off strongly resemble those created by the rigid sphere in (a), except for the presence of small-scale undulations on the cavity walls due to sphere vibration. The sphere in (c) has a shear modulus an order of magnitude smaller than that in (b); it deforms significantly upon impact creating a much wider cavity and shallower pinch-off event. The smaller $G_{\infty}$ results in higher magnitude but lower frequency oscillations, creating a second impact-like event within the first cavity. Deformations are even more pronounced in sequence (d). Pinch-off occurs within the second cavity formed for image sequences (c) and (d).  Spheres with lower values of $G_{\infty}$ are often observed to decelerate so rapidly that they occupy the space where a deep seal would normally occur as seen in Fig.~\ref{fig:allsoft_vs_rigid}e. In this instance the contact line of the second cavity recedes up the surface of the sphere and pinches-off at the top.

\begin{figure}
  \centering
  \includegraphics[width=\textwidth]{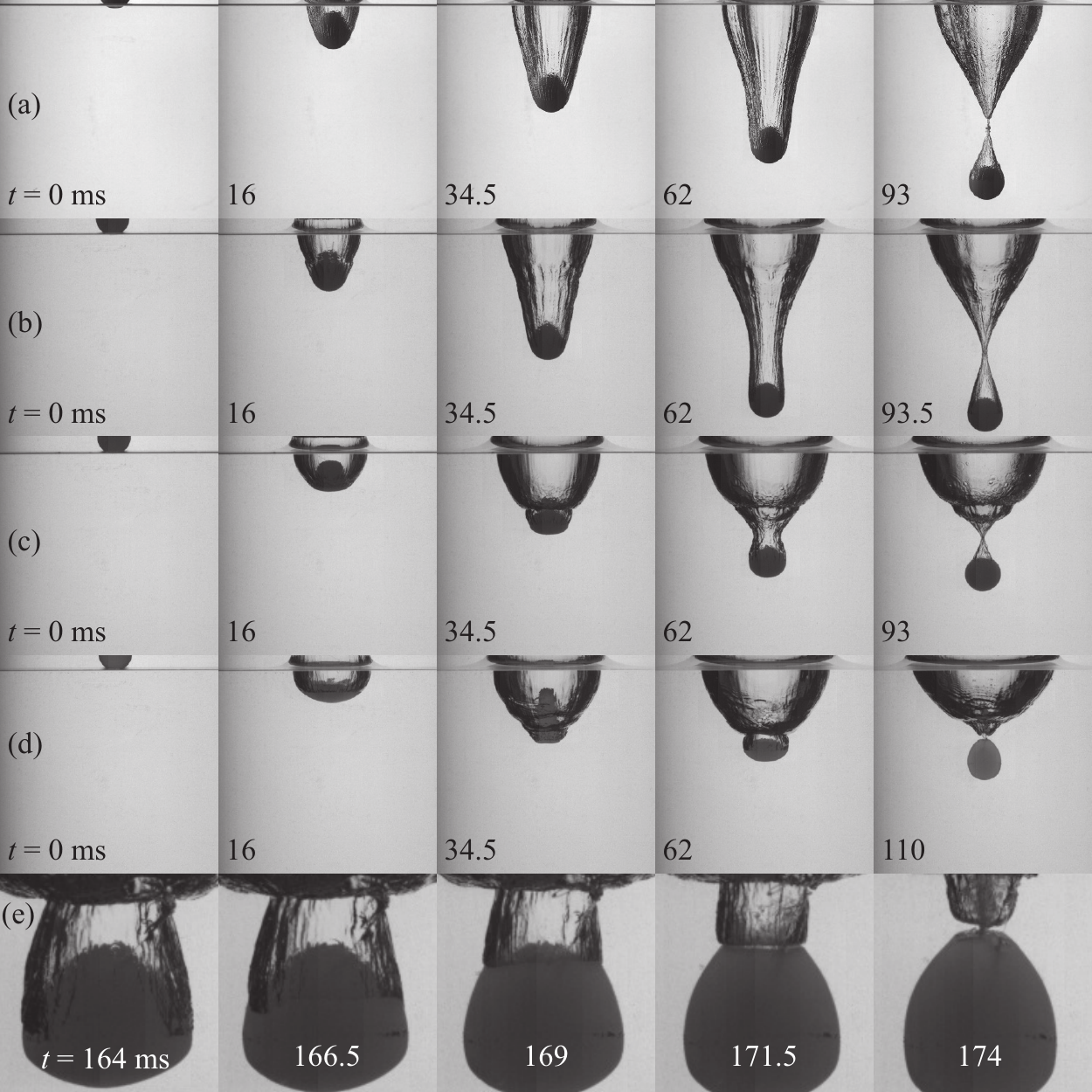}
  \caption[$soft_vs_rigid$]{ 
(a)-(d) The water entry of four spheres with identical diameter ($D$ = 51 mm), density ($\rho_s = 1070$  $\mathrm{kg/m^3}$) and impact velocity ($U_0$ = 6.5~m/s) but varying shear moduli: (a) $G_{\infty}$ = $5.66 \times 10^5$ kPa, (b) $G_{\infty}$ = 70.2 kPa, (c) $G_{\infty}$ = 6.70 kPa, (d) $G_{\infty}$ = 1.12 kPa. (e) For the largest and most compliant spheres tested ($D = 100$ mm, $U_0 = 6.5$m/s and $G_{\infty} = $ 1.12 kPa) spheres decelerate more rapidly, occupying the space where pinch-off would occur. The attached cavity recedes upward along the sphere, pinching at the top of. Image sequences (a)-(e) correspond to supplemental movies 3-7 respectively.
}
  \label{fig:allsoft_vs_rigid}
\end{figure}

Water entry events are often classified by cavity characteristics, with a common parameter being time to pinch-off ($t_p$). The dimensionless time $t_p U_0/D$ is plotted against Froude number ($Fr = U_0 / \sqrt{gD}$) in Fig.~\ref{fig:pinch_off}a for all tested cases, which is the same non-dimensionalization employed by \citet{Aristoff2010} for decelerating rigid spheres (dashed line). However, the scaling does not provide an effective data collapse for deformable elastomeric spheres. Instead we normalize using a new term $D_{\mathrm{eff}} = \lambda_{pN} D$, which represents the maximum deformed diameter that the sphere assumes within the cavity in which pinch-off occurs. For example, in the case seen in Fig.~\ref{fig:allsoft_vs_rigid}d, pinch-off occurred within the second cavity resulting in $D_{\mathrm{eff}}$ = $\lambda_{p2} D$. This adjustment provides a more convincing data collapse as can be seen in Fig.~\ref{fig:pinch_off}b, where the solid line is a fit to the data (slope = 1.3).

\begin{figure}
  \centering
  \includegraphics[width=\textwidth]{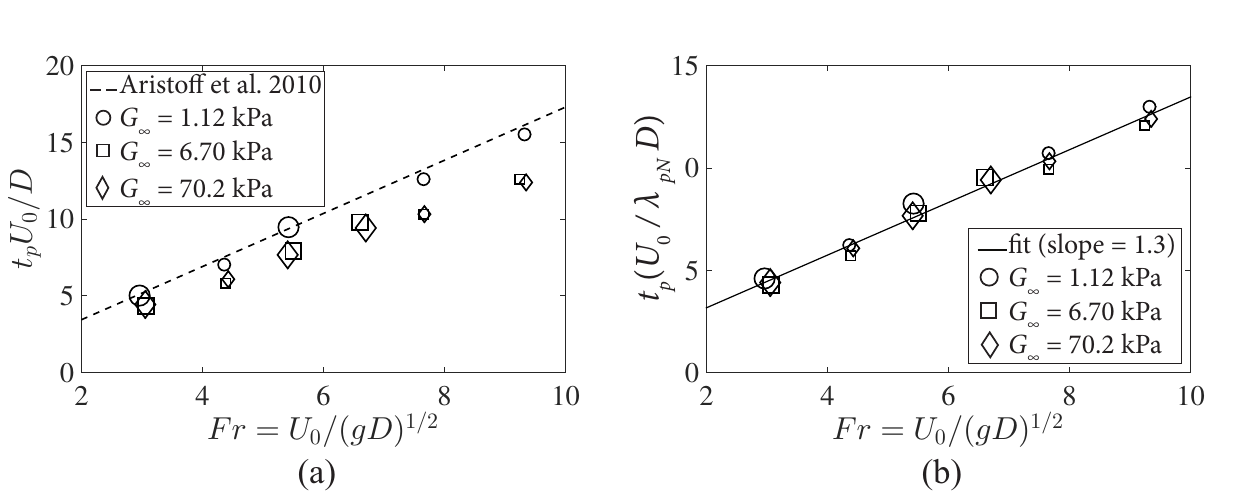}
  \caption[Pinch-off time scaling]{ 
(a) Dimensionless pinch-off plotted against Fr does not produce a convincing collapse as was the case with rigid spheres in the study by \citet{Aristoff2010}. The dashed line represents the theoretical scaling proposed in the same study. (b) Rather a more accurate scaling is achieved by a dimensionless pinch-off time scaled by $D_{\mathrm{eff}} = \lambda_{pN}$, which represents the maximum deformed diameter that the sphere assumes within the cavity in which pinch-off occurs (slope of linear fit = 1.3). }
\vspace{-0.3cm}
  \label{fig:pinch_off}
\end{figure}

We have described the effects of elastomeric sphere deformation on the global features of water entry, and now turn attention to sphere dynamics.  Based on the description of the sphere as an ellipsoid (Fig.~\ref{fig:lambda_scaling}a), the position of the center of mass is defined as
\begin{equation} \label{eq:y_c}
{y}_c = {y}_b + \frac{R}{\lambda^2},
\end{equation}
where, as already discussed, $y_b$ is tracked from images (Fig.~\ref{fig:exp}c).  Any noise in measurements of $\lambda$ would be amplified in Eq.~\ref{eq:y_c}; therefore, we use the time-scaled results of the model simulations to define $\lambda$.  The velocity and acceleration of the center of mass, $\dot{y}_c$ and $\ddot{y}_c$, are computed from derivatives of smoothing splines fit to $y_c$, as was done in~\citep{Truscott2012}.  Fig.~\ref{fig:pos_n_vel}a-c displays $\dot{y}_c$ as a function of dimensionless time for all values of shear modulus $G_{\infty}$, where $D$ = 51 mm and $U_0$ ranges from 3.0 - 6.5 m/s. The values of $\dot{y}_c$ for rigid spheres ($G_{\infty}$ =  5.66 $\times 10^5 $) are plotted as blue curves.  The vertical lines indicate the end of the first oscillation period for the elastomeric spheres (corresponding grey shades match the legend in (a)). Elastomeric spheres experience a greater deceleration than the rigid spheres as $\lambda D$ increases. However, the deformable spheres quickly transition to a deceleration rate similar to that of a rigid sphere as the magnitude of $\lambda$ decreases (compare slopes of grey curves to blue curve after the first oscillation period marked by the vertical lines). Finally, after pinch-off ($t/t_p > 1$), a steady state is reached and spheres fall at nearly constant velocity ($\lambda \rightarrow 1$). Notice that in (c) the softest spheres (lightest grey) lose nearly all of their velocity during the first deformation cycle, whereas more rigid spheres lose a significantly smaller portion.


\begin{figure}
  \centering
  \includegraphics[width=\textwidth]{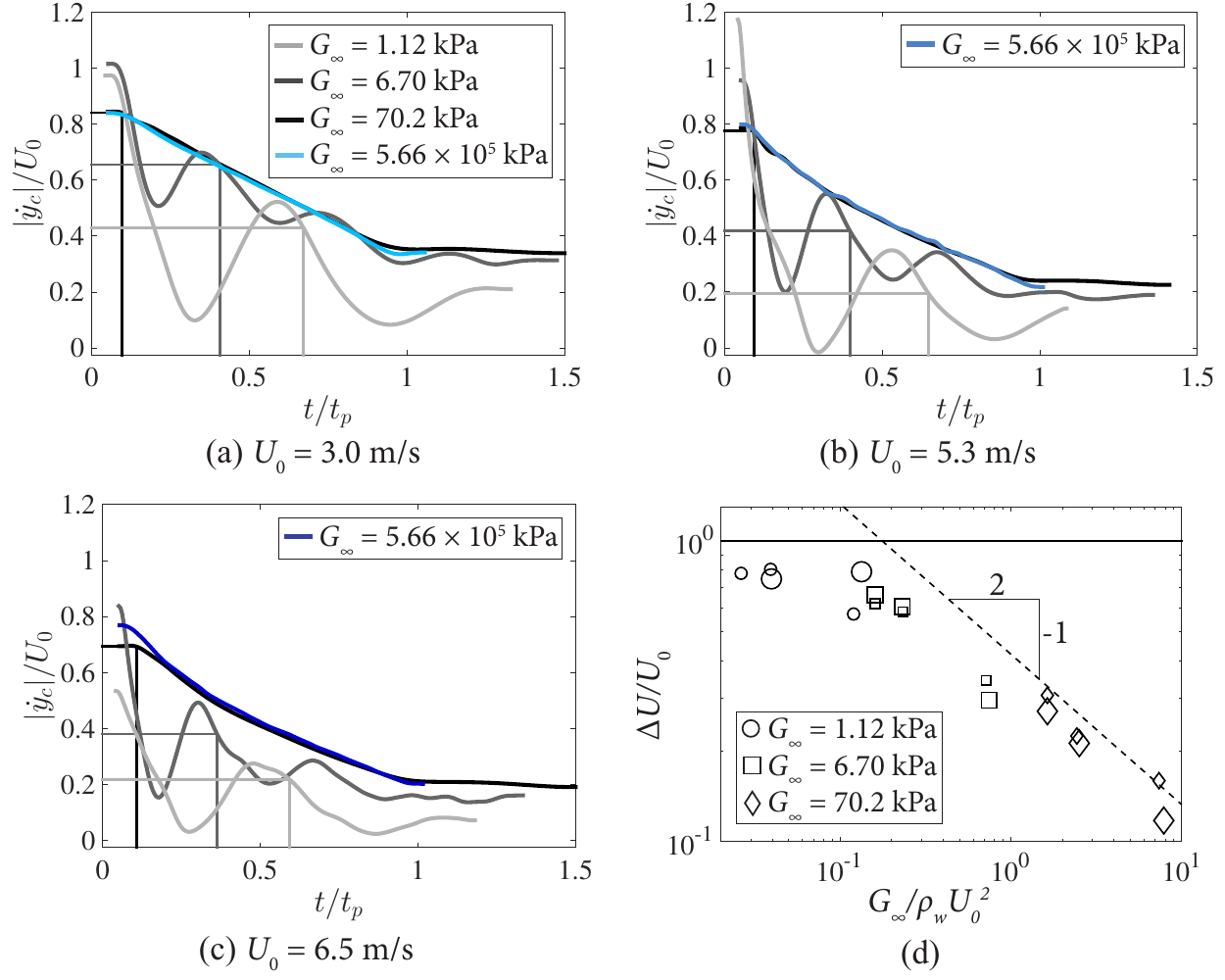}
  \caption[Position and velocity of the sphere during entry]{ 
Dimensionless velocity ($|\dot{y}_c|/U_0$) is plotted against dimensionless time ($t/t_p$) for spheres impacting with three different velocities: (a) $U_0 = 3.0$ m/s, (b) $U_0 = 5.3$ m/s and (c) $U_0 = 6.5$ m/s. Compared to rigid spheres (blue curves), deformable spheres experience a larger deceleration rate after impact over the first cycle of sphere deformation. After the first oscillation period, deformable spheres follow a deceleration similar to rigid spheres, and then transition to a nearly constant velocity after pinch-off. (d) The large initial change in velocity is investigated by plotting $\Delta U$ = $U_1 - U_0$, where $U_1$ is the velocity of the sphere center of mass after one oscillation period, against $G_{\infty}/\rho_w U_0^2$. For $G_{\infty}/\rho_w U_0^2 \gtrsim 0.2$, $\Delta U / U_0$ scales with $(G_{\infty}/\rho_w U_0^2)^{-1/2}$, as predicted by a scaling analysis from the equation of motion for the sphere. For $G_{\infty}/\rho_w U_0^2 \lesssim 0.2$, $\Delta U / U_0$ asymptotes to the limit of 1, with nearly all of $U_0$ being lost over the first period of oscillation.
}
\vspace{-0.3cm}
  \label{fig:pos_n_vel}
\end{figure}

We perform a scaling analysis of the water entry event to gain insight into the sphere deceleration over the first deformation period.  For simplicity, added mass is neglected and thus the dominant forces include drag, gravity and buoyancy. Because the spheres are nearly neutrally buoyant, gravitational and buoyant terms cancel and a simple equation of motion for the impacting sphere can be expressed as
\begin{equation} \label{eq:eom1}
\rho_s  \forall \ddot{y}_c = \frac{1}{2} \rho_w A U^2 C_D,
\end{equation}
where $\forall$ represents the volume of the sphere, $A$ the cross-sectional area, $U$ the velocity of the center of mass and $C_D$ the coefficient of drag. We simplify this expression by defining a characteristic acceleration $(U_1 - U_0)/T = \Delta U/T$, where $U_1$ denotes the velocity of the sphere center of mass after the first deformation period. By noting that $\forall  \sim D^3$, $A \sim D^2$ and $C_D \sim 1$, we can approximate Eq.~\ref{eq:eom1} as 
\begin{equation} \label{eq:eom2}
\frac{\Delta U}{T} \approx \frac{\rho_w}{\rho_s} \frac{U_0^2}{D}. 
\end{equation}
We previously showed that $T \sim D / \sqrt{G_{\infty}/\rho_s}$ (Fig.~\ref{fig:lambda_scaling}b), and for tested spheres $\rho_w / \rho_s \sim 1$. This allows us to rearrange Eq.~\ref{eq:eom2} to 
\begin{equation} \label{eq:eom3}
\frac{\Delta U}{U_0} \approx \left( \frac{G_{\infty}}{ \rho_w U_0^2} \right)^{-1/2}.
\end{equation}
The velocity $\Delta U / U_0$  is plotted against $G_{\infty}/\rho_w U_0^2$ in Fig.~\ref{fig:pos_n_vel}d.  This dimensionless number, which is a ratio of material shear modulus to impact hydrodynamic pressure, collapses the data.  For $G_{\infty}/\rho_w U_0^2 \gtrsim$ 0.2, the data follow the scaling predicted by Equation~\ref{eq:eom3}. However, in the limit of small $G_{\infty}$ and large $U_0$ spheres deform significantly, and the argument $A \sim D^2$ no longer holds as there is a more complicated dependence of $\lambda$ on the material properties and impact conditions.  Furthermore, it is likely that added mass plays a more significant role as $G_{\infty}/\rho_w U_0^2 \rightarrow 0$ (see Appendix~\ref{appB}).  When $G_{\infty}/\rho_w U_0^2 < 0.2$, we find $\Delta U \rightarrow U_0$ within the first oscillation period.  Nonetheless, the experimental data follow the proposed scaling well, and this allows us to predict how the impact dynamics of deformable spheres will differ from their rigid counterparts based on material properties and impact conditions.

At this point, it is worth commenting on the expected role of added mass during the water entry event.  Prior research on rigid sphere water entry has shown that forces arising from added mass are significant in the early moments of impact, primarily at times before the entire sphere has passed the free-surface~\citep{Faltinsen1998, Truscott2012, Truscott2014}.  As discussed earlier, it is likely that the added fluid mass is responsible for the longer oscillation period of the spheres in water.  This added mass would be expected to resist sphere acceleration in the direction of travel as the sphere oscillates.  While added mass undoubtedly affects the physics of deformable sphere water impact, we argue that it is unlikely to significantly affect the trends in $\Delta U / U_0$ for $G_{\infty}/\rho_w U_o^2 \gtrsim 0.2$ (see Appendix~\ref{appB} for more details).  This is supported by the good agreement between the experimental data and the predicted trend in Fig.~\ref{fig:pos_n_vel}d.

\begin{figure}
  \centering
  \includegraphics[width=\textwidth]{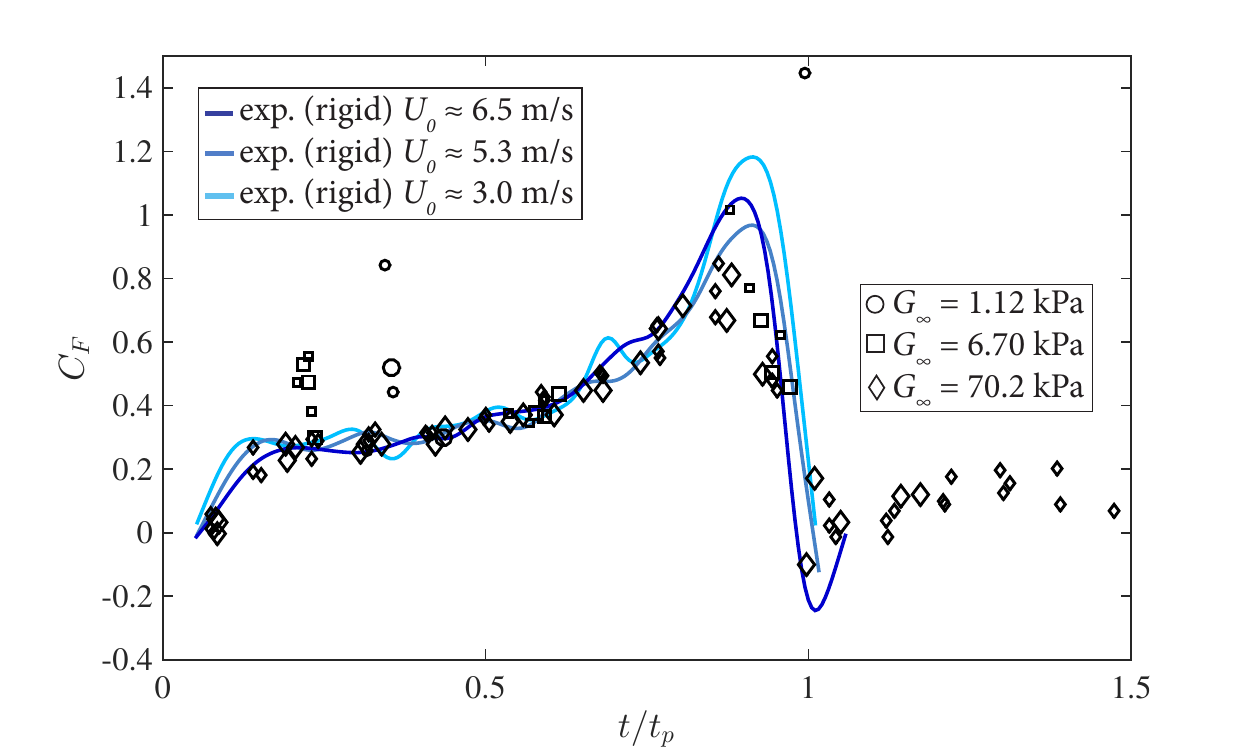}
  \caption[Drag coefficient]{ 
Force coefficients for deformable spheres ($\overline{C}_F$) calculated by averaging over each deformation period ($y$-direction) are plotted as a function of dimensionless time (black symbols). The data encompass all sphere diameters, shear moduli and impact velocities tested. Force coefficients ($C_F$) for rigid spheres entering the free surface with the same specific gravity as the deformable spheres are plotted as a function of dimensionless time (blue curves). The period averaged values for deformable spheres $\overline{C}_F$ follow the instantaneous values for rigid spheres $C_F$, except during the first sphere deformation period in which deformable spheres experience larger drag from increased $\lambda D$.}
\vspace{-0.3cm}
  \label{fig:drag_coeff}
\end{figure}

To further investigate how the water entry of a deformable elastomeric sphere differs from that of a rigid sphere, we calculate the total force coefficient acting on the sphere in the $y$-direction as a function of time. The oscillating behavior of the sphere results in a varying instantaneous force coefficient $C_F$. Therefore, we calculate a period-averaged force coefficient,
\begin{equation} \label{eq:e_drag}
\overline{C}_F = \frac{ \rho_s \forall \overline{\ddot{y}}_c }{ \frac{1}{2} \rho_w \overline{\dot{y}}_b^2 \pi \left(\frac{\lambda_{pN}D}{2}\right)^2 },
\end{equation}
where $\overline{\ddot{y}}_c$ and $\overline{\dot{y}}_b$ are the acceleration and velocity of the center of mass and sphere bottom averaged over a single oscillation period, respectively. Using Eq.~\ref{eq:e_drag}, values for $\overline{C}_F$ are plotted in Fig.~\ref{fig:drag_coeff} as a function of dimensionless time $t/t_p$ for all experimental cases. The period-averaged values $\overline{C}_F$ follow the instantaneous experimental values $C_F$ for three cases of rigid sphere water entry (blue curves). This trend holds except for the first sphere deformation period ($t/t_p  \approx$ 0.2 to 0.4 depending on $G_{\infty}$), for which the deformable spheres experience larger drag from increased $\lambda D$.  Over this period, the spheres deform into ellipsoids with a large aspect ratio and thus we expect the force coefficient to be larger.  For example, for an ellipsoid with $\lambda$ = 1.3, we expect the force coefficient to be between 3-7 times larger than that of a sphere, depending on Reynolds number \citep{Daugherty1977}.

\section{Conclusion}
We have shown that deformable elastomeric spheres form cavities that differ from those formed by rigid spheres by being shallower, wider and having undulatory cavity walls. These differences stem from the sphere flattening upon surface impact followed by material oscillation. We describe the deformation and oscillation in terms of both material properties and impact conditions. This allows us to define an effective diameter, which accounts for the deformation and provides effective scaling for time to pinch-off and a period-averaged force coefficient. The large sphere deformation, particularly over the first period, is responsible for the increased loss in velocity as compared to rigid spheres.  We have shown how this reduction in velocity scales with the ratio of material shear modulus to impact hydrodynamic pressure ($G_{\infty}/\rho_w U_0^2$). 
Surprisingly, we find that except for the unique initial deceleration and altered cavity dynamics, which we have quantified in terms of stiffness and impact velocity, the dynamics for the water entry of deformable elastomeric spheres mirror that of rigid spheres.

\section*{Acknowledgements}
J.B., T.T.T. and R.C.H. acknowledge funding from the Office of Naval Research, Navy Undersea Research Program (grant N0001414WX00811), monitored by Ms. Maria Medeiros. J.B. and M.A.J. acknowledge funding from the Naval Undersea Warfare Center In-House Lab Independent Research program, monitored by Mr. Neil Dubois. 

\bibliographystyle{jfm}
\bibliography{waboba_normal}

\appendix
\section{Viscoelastic model}
\label{appA}

\subsection{Describing the sphere deformation}
\label{sec:deformation}

The model of sphere deformation is shown in Fig.~\ref{fig:sphere_deformation}.  The deformation is described by assuming a volume preserving stretch that deforms the sphere into an ellipsoid, with semi axes $\left( \lambda R, \lambda R, \lambda_3 R \right)$ aligned with the $\mathbf{e_1}-\mathbf{e_2}-\mathbf{e_3}$ coordinate system.  The incompressibility condition requires that $\lambda_3 = 1/\lambda^2$.  The total deformation gradient can be expressed as
\begin{equation}
\mathbf{F} = \lambda \left( \mathbf{e_1} \otimes \mathbf{e_1} + \mathbf{e_2} \otimes \mathbf{e_2} \right) + \frac{1}{\lambda^2} \mathbf{e_3} \otimes \mathbf{e_3},
\end{equation}
where $\otimes$ denotes the tensor product of two vectors.

We suppose that the solid can be idealized as a linear viscoelastic Bergstrom-Boyce material~\citep{Bergstrom1998}.  In this model, the total deformation gradient is decomposed into elastic and plastic parts $\mathbf{F} = \mathbf{F}^e \mathbf{F}^p$.   For the simple deformation here, both $\mathbf{F}^e$ and $\mathbf{F}^p$ are volume preserving stretches parallel to the basis vectors, so we can write
\begin{eqnarray}
\mathbf{F}^p &=& \lambda_p \left( \mathbf{e_1} \otimes \mathbf{e_1} + \mathbf{e_2} \otimes \mathbf{e_2} \right) + \frac{1}{\lambda_p^2} \mathbf{e_3} \otimes \mathbf{e_3} \\
\label{eq:deformation_gradient_decomp_p}
\mathbf{F}^e &=& \lambda_e \left( \mathbf{e_1} \otimes \mathbf{e_1} + \mathbf{e_2} \otimes \mathbf{e_2} \right) + \frac{1}{\lambda_e^2} \mathbf{e_3} \otimes \mathbf{e_3},
\label{eq:deformation_gradient_decomp_e}
\end{eqnarray}
where 
\begin{equation}
\lambda = \lambda_e \lambda_p.
\label{eq:lambda_relations}
\end{equation}
This allows us to calculate the Left Cauchy-Green deformation tensor for the total and elastic deformation gradients
\begin{eqnarray}
\mathbf{B} &=& \mathbf{F} = \lambda^2 \left( \mathbf{e_1} \otimes \mathbf{e_1} + \mathbf{e_2} \otimes \mathbf{e_2} \right) + \frac{1}{\lambda^4} \mathbf{e_3} \otimes \mathbf{e_3} 
\label{eq:CauchyGreen} \\
\mathbf{B}^e &=& \mathbf{F} \mathbf{F}^T = \lambda_e^2 \left( \mathbf{e_1} \otimes \mathbf{e_1} + \mathbf{e_2} \otimes \mathbf{e_2} \right) + \frac{1}{\lambda_e^4} \mathbf{e_3} \otimes \mathbf{e_3}.
\label{eq:CauchyGreen_e}
\end{eqnarray}

\begin{figure} 
\centering
\includegraphics[width=0.5\textwidth]{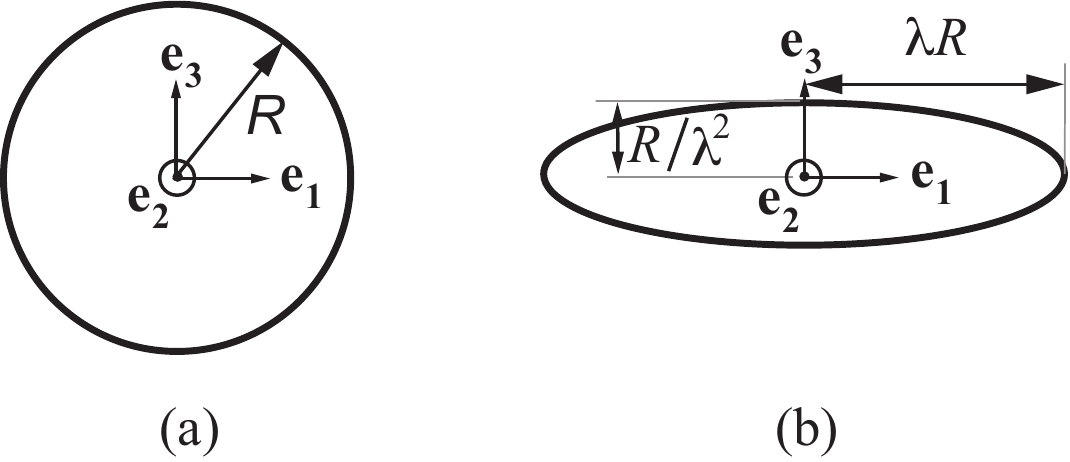}
\caption{Nominal deformation of the sphere into an axisymmetric ellipsoid.}
\label{fig:sphere_deformation}
\end{figure}
 
\noindent The invariants of the tensors are
\begin{eqnarray}
I_1 &=& \mathrm{tr}(\mathbf{B}) = 2 \lambda^2 + \frac{1}{\lambda^4} \\
\label{eq:I1}
I_2 &=& \frac{1}{2}(I_1^2 - \mathbf{B}:\mathbf{B}) = \frac{1}{2} \left[ \left(2\lambda^2 + \frac{1}{\lambda^4} \right)^2 - \left(2\lambda^4 + \frac{1}{\lambda^8} \right)\right] \\
\label{eq:I2}
I_1^e &=& \mathrm{tr}(\mathbf{B}^e) = 2 \lambda_e^2 + \frac{1}{\lambda_e^4} \\
\label{eq:I1_e}
I_2^e &=& \frac{1}{2}({I_1^e}^2 - \mathbf{B}:\mathbf{B}) = \frac{1}{2} \left[ \left(2\lambda_e^2 + \frac{1}{\lambda_e^4} \right)^2 - \left(2\lambda_e^4 + \frac{1}{\lambda_e^8} \right)\right]. 
\label{eq:I2_e}
\end{eqnarray}
We also need measures of total, elastic and plastic strain rates.   We use the symmetric part of the velocity gradient as the strain rate measure 
\begin{eqnarray}
\mathbf{D} &=& \mathrm{sym} \left( \dot{\mathbf{F}} \mathbf{F}^{-1} \right) = \mathbf{D}^e + \mathbf{D}^p \\
\label{eq:D}
\mathbf{D}^e &=& \mathrm{sym} \left( \dot{\mathbf{F}^e} {\mathbf{F}^e}^{-1} \right)  \\
\label{eq:D_e}
\mathbf{D}^p &=& \mathrm{sym} \left( \mathbf{F}^e \dot{\mathbf{F}^p} {\mathbf{F}^p}^{-1} {\mathbf{F}^e}^{-1} \right).
\end{eqnarray}
For the simple stretch considered here, we get
\begin{equation}
\mathbf{D}^p = \frac{\dot{\lambda_p}}{\lambda_p} \left( \mathbf{e_1} \otimes \mathbf{e_1} + \mathbf{e_2} \otimes \mathbf{e_2} \right) - 2\frac{\dot{\lambda_p}}{\lambda_p} \mathbf{e_3} \otimes \mathbf{e_3}.
\label{eq:deformation_gradient}
\end{equation}

\subsection{Material model theory}
\label{sec:material_model_theory}

For the special case of an incompressible material, the Bergstrom-Boyce model assumes that the stress can be derived from an elastic strain energy of the form
\begin{equation}
U \left( I_1, I_2, I_1^e, I_2^e \right) = U_{\infty} \left( I_1, I_2 \right) + U_T \left(I_1^e, I_2^e \right).
\label{eq:strain_energy}
\end{equation}
We can regard this as a nonlinear version of the 3-parameter Maxwell model (Fig.~\ref{fig:element}), in which $U_T$ represents the energy in spring $k_1$ (this energy eventually relaxes to zero if a constant strain is applied to the material) and $U_{\infty}$ represents the energy in spring $k_2$.

The stresses are related to the derivatives of the strain energy in the usual way, giving
\begin{equation}
\mathbf{\sigma} = 2 \left[ \left( \frac{\partial U_{\infty}}{\partial I_1} + I_1 \frac{\partial U_{\infty}}{\partial I_2} \right) \mathbf{B} - \frac{\partial U_{\infty}}{\partial I_2} \mathbf{B}^2 \right] + 2 \left[ \left( \frac{\partial U_{T}}{\partial I_1^e} + I_1^e \frac{\partial U_{T}}{\partial I_2^e} \right) \mathbf{B}^e - \frac{\partial U_{T}}{\partial I_2^e} \mathbf{B}^{e^2} \right] + p \mathbf{1}.
\label{eq:stress_strain}
\end{equation}

To model the material used for the spheres presented in this paper, we choose $U_{\infty}$ and $U_{T}$ to be the incompressible Neo-Hookean potential
\begin{eqnarray}
U_{\infty} &=& \frac{G_{\infty}}{2} \left( I_1 - 3 \right) \\
\label{eq:U_inf}
U_{T} &=& \frac{G_{T}}{2} \left( I_1^e - 3 \right).
\label{eq:U_T}
\end{eqnarray}
For this choice, we get
\begin{equation}
\mathbf{\sigma} =  G_{\infty} \mathbf{B} + G_{T} \mathbf{B}^e + p \mathbf{1}.
\label{eq:stress_strain_2}
\end{equation}
We also need an evolution equation for the plastic part of the stretch $\mathbf{F}^p$.  Bergstrom-Boyce suggest the following equation:
\begin{equation}
\mathbf{D}^p =  \dot{\epsilon}_0 \left(\sqrt{I_1^p} -\sqrt{3} + \xi \right)^n \left( \frac{\tau_e}{\tau_0} \right)^m \frac{3}{2} \frac{\boldsymbol{\tau}}{\tau_e},
\label{eq:D_p}
\end{equation}
where $\dot{\epsilon}_0$, $m$, $n$, $\tau_0$ are material properties, $\xi$ is a constant, $\mathbf{\tau} = G_T \left(\mathbf{B}^e - \frac{1}{3} tr\left( \mathbf{B}^e \right) \mathbf{1} \right)$ is the deviatoric part of the `dynamic' stress, and $\tau_e = \sqrt{3 \boldsymbol{\tau}:\boldsymbol{\tau}/2}$ is the Von Mises uniaxial equivalent dynamic stress.  For the volume preserving stretching deformation considered here,
\newpage
\begin{eqnarray}
\frac{\boldsymbol{\tau}}{G_T} &=& \lambda_e^2 \left( \mathbf{e_1} \otimes \mathbf{e_1} + \mathbf{e_2} \otimes \mathbf{e_2} \right)  + \frac{1}{\lambda_e^4} \mathbf{e_3} \otimes \mathbf{e_3} - \frac{1}{3} \left( 2\lambda_e^2 + \frac{1}{\lambda_e^4} \right) \left( \mathbf{e_1} \otimes \mathbf{e_1} + \mathbf{e_2} \otimes \mathbf{e_2} + \mathbf{e_3} \otimes \mathbf{e_3} \right) \nonumber \\
 &=& \frac{1}{3} \left( \lambda_e^2 - \frac{1}{\lambda_e^4} \right) \left( \mathbf{e_1} \otimes \mathbf{e_1} + \mathbf{e_2} \otimes \mathbf{e_2} - 2\mathbf{e_3} \otimes \mathbf{e_3} \right),
\label{eq:tau}
\end{eqnarray}
and
\begin{equation}
\tau_e=  G_{T} \left| \lambda_e^2 - \frac{1}{\lambda_e^4}  \right|.
\label{eq:stress_strain_rch}
\end{equation}
Invoking Eq.~\ref{eq:deformation_gradient} in Eq.~\ref{eq:D_p} we get
\begin{equation}
\frac{\dot{\lambda}_p}{\lambda_p} =  \frac{1}{2} \dot{\epsilon}_0 \left(\sqrt{I_1^p} -\sqrt{3} + \xi \right)^n \left( \frac{\tau_e}{\tau_0} \right)^m \mathrm{sign} \left( \lambda_e^2 - \frac{1}{\lambda_e^4} \right),
\label{eq:lambda_p}
\end{equation}
where
\begin{equation}
I_1^p = 2 \lambda_p^2 +\frac{1}{\lambda_p^4}
\label{eq:I_1_p}
\end{equation}

\subsection{Dynamics}
\label{sec:dynamics}

\begin{figure} 
\centering
\includegraphics[width=0.4\textwidth]{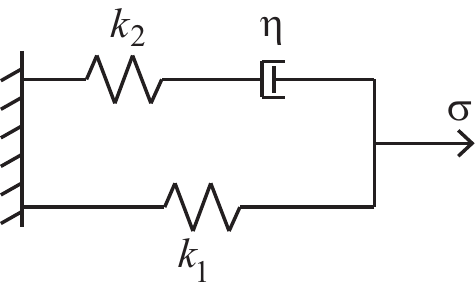}
\caption{Nonlinear version of the 3-parameter Maxwell model that provides the framework for describing the stress in the Bergstrom-Boyce model.}
\label{fig:element}
\end{figure}

Finally, we need the equation of motion for $\lambda$, which will be obtained from the principle of virtual work~\cite{Bower2009}
\begin{equation}
\int\limits_V (\boldsymbol{\sigma}:\delta \mathbf{D}) dV + \int\limits_V \rho_{\mathrm{s}} (\mathbf{a} \cdot \mathbf{\delta v}) dV + \int\limits_V \rho_{\mathrm{s}} (\mathbf{b} \cdot \mathbf{\delta v}) dV - \int\limits_A (\mathbf{t} \cdot \mathbf{\delta v}) dA = 0, 
\label{eq:virtual_work_1a}
\end{equation}
where $\delta \mathbf{v} = \delta\mathbf{\dot{V}}\mathbf{x}$ is a virtual velocity field, $\mathbf{x}$ denote the coordinates of a material particle before deformation and  
\begin{equation}
\delta\mathbf{\dot{V}} = \delta \dot{\lambda} \left( \mathbf{e_1} \otimes \mathbf{e_1} + \mathbf{e_2} \otimes \mathbf{e_2} \right) - \frac{2 \delta \dot{\lambda}}{\lambda} \mathbf{e_3} \otimes \mathbf{e_3}. 
\label{eq:virtual_stretch}
\end{equation}
In this analysis, we neglect the effects of gravity and assume there are no external tractions; thus the third and fourth terms in Eq.~\ref{eq:virtual_work_1a} vanish.    Invoking Eqs.~\ref{eq:CauchyGreen}-\ref{eq:CauchyGreen_e} \&~\ref{eq:stress_strain_2}, the first term in Eq.~\ref{eq:virtual_work_1a} becomes
\begin{equation}
\int\limits_V (\boldsymbol{\sigma}:\delta \mathbf{D}) dV = \frac{4}{3} \pi R^3 \left[ 2 G_{\infty} \left( \lambda - \frac{1}{\lambda^5} \right) + 2G_T \frac{1}{\lambda} \left( \lambda_e^2 - \frac{1}{\lambda_e^4} \right) \right] \delta \dot{\lambda}.
\label{eq:virtual_work_term1}
\end{equation}
\newpage
To evaluate the remaining terms, the following identities are useful
\begin{eqnarray}
\int\limits_{V_0} dV_0 &=& \frac{4\pi}{3}R^3 \nonumber \\ 
\int\limits_{V_0} x_i dV_0 &=& 0 \nonumber \\ 
\int\limits_{V_0} x_i x_j dV_0 &=& \frac{4\pi}{15}R^5 \delta_{ij}
\label{eq:identities_2} 
\end{eqnarray} 
where $x_i$ denote the coordinates of a material particle with respect to the center of the sphere and the integrals are evaluated over the undeformed sphere. The inertia term (second term in Eq.~\ref{eq:virtual_work_1a}) can be expressed as 
\begin{eqnarray}
\int\limits_V \rho_{\mathrm{s}} (\mathbf{a} \cdot \mathbf{\delta v}) dV  =  \int\limits_{V} \rho_{\mathrm{s}} \left[ \mathbf{\ddot{F}}\mathbf{x}  \right] \cdot \left[ \delta\mathbf{\dot{V}}\mathbf{x} \right] dV &=& \frac{4 \pi}{15} \rho_{\mathrm{s}} R^5 \left[ \mathbf{\ddot{F}} \mathbf{F} : \delta\mathbf{\dot{V}} \mathbf{F}^{-T} \right]
 \nonumber \\
 &=& 2 \left( \ddot{\lambda} - \frac{6 \dot{\lambda}^2 - 2 \lambda \ddot{\lambda}}{\lambda^7} \right) \delta \dot{\lambda}.
\label{eq:virtual_work_term2}
\end{eqnarray}
Collecting terms gives
\begin{equation}
\frac{4}{3} \pi R^3 \left[ 2 G_{\infty} \left( \lambda - \frac{1}{\lambda^5} \right) + 2G_T \frac{1}{\lambda} \left( \lambda_e^2 - \frac{1}{\lambda_e^4} \right) \right] + \frac{8 \pi}{15} \rho_{\mathrm{s}} R^5 \left( \ddot{\lambda} - \frac{6 \dot{\lambda}^2 - 2 \lambda \ddot{\lambda}}{\lambda^7} \right) = 0. 
\label{eq:eom1_appA}
\end{equation}
Equations~\ref{eq:lambda_relations}, \ref{eq:stress_strain_rch}-\ref{eq:I_1_p} \&~\ref{eq:eom1_appA} are solved in Matlab to resolve the stretch $\lambda(t)$ given initial conditions $\lambda(0) = \lambda_0$,  $\lambda_p(0) = 1$ and $\dot{\lambda} = \dot{\lambda_e} = \dot{\lambda_p} = 0$.

\subsection{Material model calibration}
\label{sec:material_model_calibration}

The material model is defined by 7 parameters: $G_{\infty}$, $G_T$, $\dot{\epsilon}_0$, $m$, $n$, $\tau_0$ and $\xi$.  Without access to the material testing facilities that would be required to fully characterize the silicone materials used in this paper, we adopt a two-part approach to estimate parameters.  First, the long time modulus $G_{\infty}$ is estimated from quasi-static testing in which the actual spheres used in the water entry experiments are compressed on an Instron machine. 

This test setup was then numerically modeled using the finite element software Abaqus where the sphere was modeled as an axisymmetric solid compressed between two rigid planes accounting for large deformation and frictionless contact. Commanding a displacement profile to match the experimental values, the resulting force is observed.  Minimizing the difference in force between the numerical and experimental results is achieved by varying the neo-Hookean shear modulus, $G_{\infty}$.  The assumption here is that the response is slow enough that the behavior is quasi-static and all rate effects can be neglected, thus we only need to calibrate one parameter \citep{Abaqus2016}. 
This is consistent with the strain energy defined in Eq.~\ref{eq:U_inf}.  We then varied $G_{\infty}$ to find the value that produced the best fit between the numerically simulated and experimentally measured force-displacement curves.  The results of these tests and numerical simulations are shown in Fig.~\ref{fig:neo_Hookean_calib}.

\begin{figure} 
\centering
\includegraphics[width=1\textwidth]{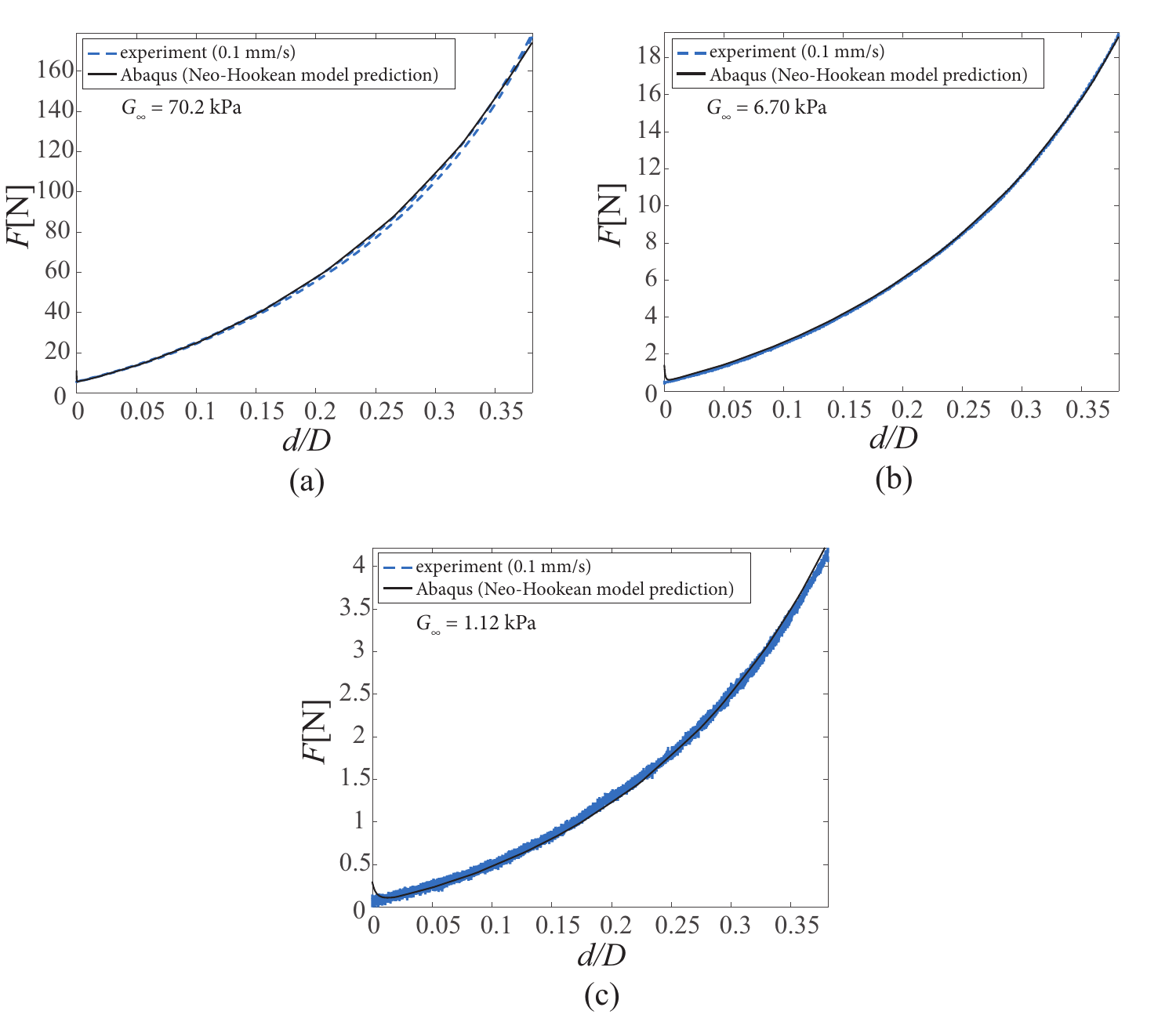}
\caption{Results from quasi-static testing in which the actual spheres used in the water entry experiments are compressed on an Instron machine at a rate of 0.1 mm/s.  The dashed blue lines show the experiments measurements of force as a function of normalized displacement $d/D$.  The solid black lines are predictions from an Abaqus simulation using a hyperelastic neo-Hookean model.  The shear modulus $G_{\infty}$ was adjusted to find the best fit between the simulation and experiment.  The three plots correspond to the three stiffness values: (a) $G_{\infty} = 70.2$ kPa, (b) $G_{\infty} = 6.70$ kPa and (c) $G_{\infty} = 1.12$ kPa}
\label{fig:neo_Hookean_calib}
\end{figure}

To estimate the `dynamic' parameters of the material model, we perform an experiment in which all 6 spheres used in the water entry tests are dropped from 3 heights each onto a rigid horizontal surface.  The maximum stretch in the plane of the image is measured, as shown in Fig.~\ref{fig:table_impact}.  The sphere response is then simulated using the dynamic model defined in Eqs.~\ref{eq:lambda_relations}, \ref{eq:stress_strain_rch}-\ref{eq:I_1_p} \&~\ref{eq:eom1_appA} with the initial stretch $\lambda_0$ set to the peak value measured in the experiment.  We allow two parameters of the material model to be free - $\dot{\epsilon}_0$, $n$ - and perform a nonlinear least-squares minimization to find the parameters that yield the best fit to the sphere stretch measurements.  The material model parameters are summarized in Table~\ref{table:mat_params}.  The simulation results using these material parameters to model the sphere response following impact with the rigid surface are shown in Fig.~\ref{fig:table_impact}.  In modeling the sphere response during water entry, these material parameters are used and the simulations is initialized with $\lambda_0$ measured from the experiments.
\begin{table}
\small
\caption{Summary of material properties for the silicone spheres studied herein.} 
\centering 
\begin{tabular}{c c c c c c c c} 
\hline\hline 
Sphere radius, $R$ (m) & $G_\infty$ (Pa)  & $G_T$ (Pa) & $\xi$ & $\tau_0$ & $m$ & $n$ & $\dot{\epsilon}_0$ \\
\hline 
0.025 & 74690 & 74690 & 0.0866 & 1.0  & 1.0  & -0.2481 & 0.0049 \\
0.025 & 6900 & 6900  & 0.0866   & 1.0  & 1.0 & -0.1902 & 0.021  \\
0.025 & 1235 & 1235 & 0.0866   & 1.0  & 1.0 & -1.0 & 0.0056 \\
0.0487 & 74690 & 74690 & 0.0866 & 1.0  & 1.0  & -0.2402 & 0.0024 \\
0.0487 & 6900 & 6900  & 0.0866   & 1.0  & 1.0 & -0.488 & 0.0041  \\
0.0487 & 1235 & 1235 & 0.0866   & 1.0  & 1.0 & -0.50 & 0.0066 \\
\hline 
\label{table:mat_params} 
\end{tabular}
\end{table}

\begin{figure} 
\centering
\includegraphics[width=0.9\textwidth]{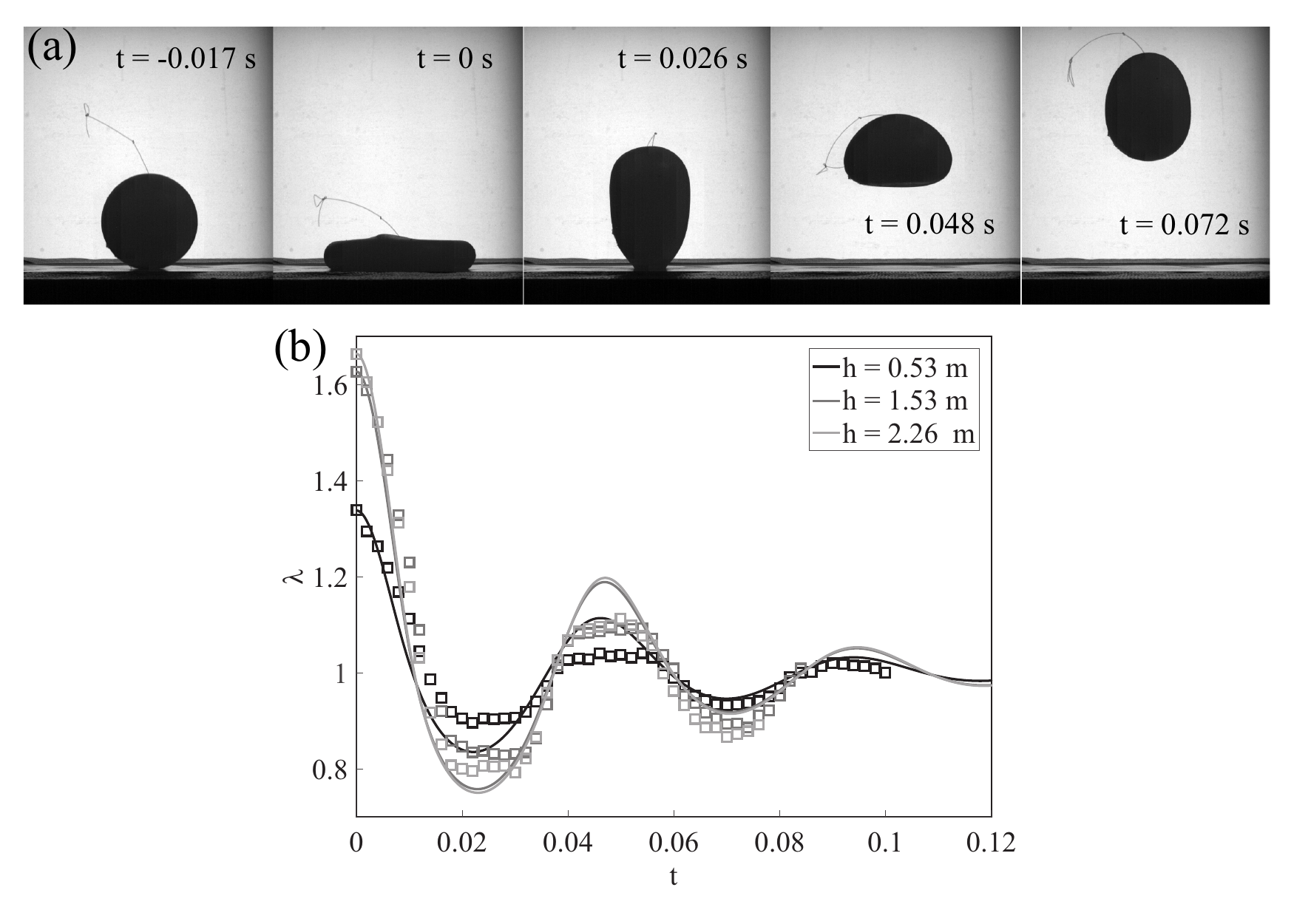}
\caption{Experiments of the spheres impacting with a rigid horizontal surface were used to calibrate the viscoelastic material model.  Results are shown for a sphere with diameter $D = $100 mm and $G_{\infty} = 6.70$ kPa.  (a) Sample high speed images from the $h = $1.53 m test.  (b) The sphere is dropped from three heights above the table and the stretch $\lambda$ is measured from high speed images (squares).  The viscoelastic parameters are found by simultaneously finding the best fit between simulations (lines) of $\lambda(t)$ and the experimental data for all drop heights for a given sphere. Shading indicates the height the spheres were dropped from.}
\label{fig:table_impact}
\end{figure}

}

\section{Added mass}\label{appB}

The scaling analysis outlined in Eqs.~\ref{eq:eom1}-\ref{eq:eom3} neglected the effect of added mass.  Here we include an added mass term in the equation of motion for $y_c$ to evaluate the affect on the scaling arguments.  Equation~\ref{eq:eom1} becomes  
\begin{equation} \label{eq:eom1_addedmass1}
\rho_s  \forall \ddot{y}_c = -\rho_w \forall C_m \ddot{y}_c + \frac{1}{2} \rho_w U^2 C_D A,
\end{equation}
where $C_m$ is an added mass coefficient.  Solving for $\ddot{y_c}$ gives
\begin{equation} \label{eq:eom1_addedmass2}
    \ddot{y_c} = \frac{\rho_w}{2 \forall \left(\rho_s + C_m \rho_w \right)} U^2 C_D A.
\end{equation}
Using the same scales as used in deriving Eq.~\ref{eq:eom3} gives
\begin{equation} \label{eq:eom1_addedmass3}
    \frac{\Delta U}{U_0} \approx \frac{\rho_w}{\rho_s + C_m \rho_w} \frac{U_0}{\sqrt{G_{\infty}/\rho_s}},
\end{equation}
and since for the spheres studied herein $\rho_s \approx \rho_w$, we find
\begin{equation} \label{eq:eom1_addedmass4}
    \frac{\Delta U}{U_0} \approx \frac{1}{1+C_m} \left( \frac{G_{\infty}}{\rho_w U_0^2} \right)^{-1/2},
\end{equation}
which differs from Eq.~\ref{eq:eom3} only by the pre-factor.  Based on values of $C_m$ for fully submerged ellipsoids~\citep{Newman1977}, we estimate a representative range of this prefactor as 0.35-0.86 corresponding a range of $\lambda = 1.5$-$0.74$.  Therefore, as $\lambda$ becomes larger, which occurs as $G_{\infty}/\rho_w U_0^2$ gets small, the added mass has a more profound affect on the relationship between $\Delta U/U_0$ and $G_{\infty}/\rho_w U_0^2$.  However, for larger values of $G_{\infty}/\rho_w U_0^2$, Eq.~ \ref{eq:eom1_addedmass4} approaches Eq.~\ref{eq:eom3}.  Indeed, for $G_{\infty}/\rho_w U_0^2 \gtrsim 0.2$ the data in Fig.~\ref{fig:pos_n_vel}d follow the trend predicted by the scaling analysis that excludes added mass.     


\end{document}